\documentclass[aps,prl,twocolumn,longbibliography]{revtex4-2}

\usepackage{amssymb}
\usepackage{amsbsy}
\usepackage{amsmath}
\usepackage{graphicx}
\graphicspath{{./}{figs/}}
\usepackage{graphics}
\usepackage{setspace}
\usepackage{array}
\usepackage{color}
\usepackage{fontenc}
\usepackage{textcomp}
\usepackage{bm}
\usepackage{float}
\usepackage[bookmarks=false,linkcolor=blue,urlcolor=blue,colorlinks,citecolor=blue]{hyperref}

\newcommand{\blue}[1]{{\color{blue}{#1}}}

\newcommand{\bsub}{\begin{subequations}}
\newcommand{\esub}{\end{subequations}}

\graphicspath{{../Figures/}}

\begin{document}
\title{Defect-Controlled Multiferroicity via Stacking Control in Nonmagnetic van der Waals Bilayers}

\author{Bumseop Kim}
\affiliation{Department of Chemistry, University of Pennsylvania, Philadelphia, Pennsylvania 19104, USA}

\author{Sayed Ali Akbar Ghorashi}\email{ghorashi@sas.upenn.edu}
\affiliation{Department of Chemistry, University of Pennsylvania, Philadelphia, Pennsylvania 19104, USA}

\author{Andrew M. Rappe}
\email{rappe@sas.upenn.edu}
\affiliation{Department of Chemistry, University of Pennsylvania, Philadelphia, Pennsylvania 19104, USA}

\begin{abstract}
We present a general paradigm that directly couples vacancy-localized magnetism to interfacial sliding ferroelectricity in nonmagnetic van der Waals (vdWs) bilayers. Utilizing bilayer hexagonal boron nitride (hBN) as a prototypical vdWs host system, 
we use first-principles calculations to show that a single vacancy acts as a local registry sensor, lifting the degeneracy between polar sliding partners via a defect-centered polarization offset. For interlayer vacancy pairs, we discover a defect selectivity where the hosting sublattice fully dictates the interlayer exchange, stabilizing either ferrimagnetic or antiferromagnetic configurations. Applying an out-of-plane electric field selects the polar registry and drives an amplitude modulation of the compensated N\'eel order parameter. These findings establish a robust, sublattice-dependent engineering of multiferroic functionality via stacking control across a wide class of nonmagnetic 2D heterostructures.    
%
\end{abstract}
\maketitle

\blue{\emph{Introduction}}.---Controlling magnetism with an electric field has long been a central goal in magnetoelectric multiferroic research because of its potential applications in low-power nonvolatile memory and spintronic devices~\cite{Spaldin2005Science,Eerenstein2006Nature,Cheong2007NatMater,Fiebig2016NatRevMater}. Realizing this functionality in the atomically thin limit is particularly attractive because vdWs materials can be electrostatically gated, vertically stacked, and integrated into nanoscale heterostructures with a degree of structural precision that is difficult to achieve in bulk materials~\cite{Geim2013Nature,Novoselov2016Science,Huang2018NatNanotech,Deng2018Nature,Fei2018Nature}. However, realizing multiferroic behavior in two dimensions (2D) remains challenging. The microscopic conditions required to establish stable electric polarization and magnetic order simultaneously are generally restrictive, and single-phase 2D materials that robustly support both orders are rare~\cite{Gong2019NatCommun,Gao2021Nanoscale,Wu2024NatCommun,Tang2025ChemMater}. 
A complementary strategy is therefore needed in which simpler microscopic elements that are individually stable and electrically controllable are combined to realize multiferroic-like functionality.

In this work, we show how sliding ferroelectricity can be converted into a local magnetic control parameter by introducing vacancy-localized magnetic moments in an otherwise nonmagnetic vdW host. 
We demonstrate the effectiveness of this approach in bilayer hBN, which serves as an ideal platform for this purpose: bilayer hBN provides both a complementary electric degree of freedom through sliding ferroelectricity and localized in-gap states and magnetic moments whose properties depend on the sublattice of the defect~\cite{SiXue2007,Yasuda2021,HuangLee2012,YangKimHongQian2010,Ouyang2013,Weston2018PRB,Sajid2020,Gottscholl2020NatMater,Stern2022,Mathur2022,Guo2023,Udvarhelyi2023,Durand2023PRL,Sortino2024,ViznerStern2021,WuLi2021,woods2021charge}. Unlike conventional ferroelectricity, which mainly arises from internal ionic distortions, the polarization in bilayer hBN is determined by interlayer sliding~\cite{WuLi2021}. Therefore, switching the registry of defective bilayer hBN from AB to BA changes not only the sign of the sliding polarization, $P_s$, but also the vertical B-N environment experienced by the vacancy. We then show that a magnetic defect embedded in bilayer hBN can couple to the sliding-ferroelectric coordinate through its stacking-dependent local environment. The B and N sublattices provide chemically different vacancy centers with controllable spin moments and energy-level alignments. 
Moreover, the wide band gap in hBN isolates these defect states from the bulk bands.

\begin{figure*}[ht]
  \centering
  \includegraphics[width=0.7\textwidth]{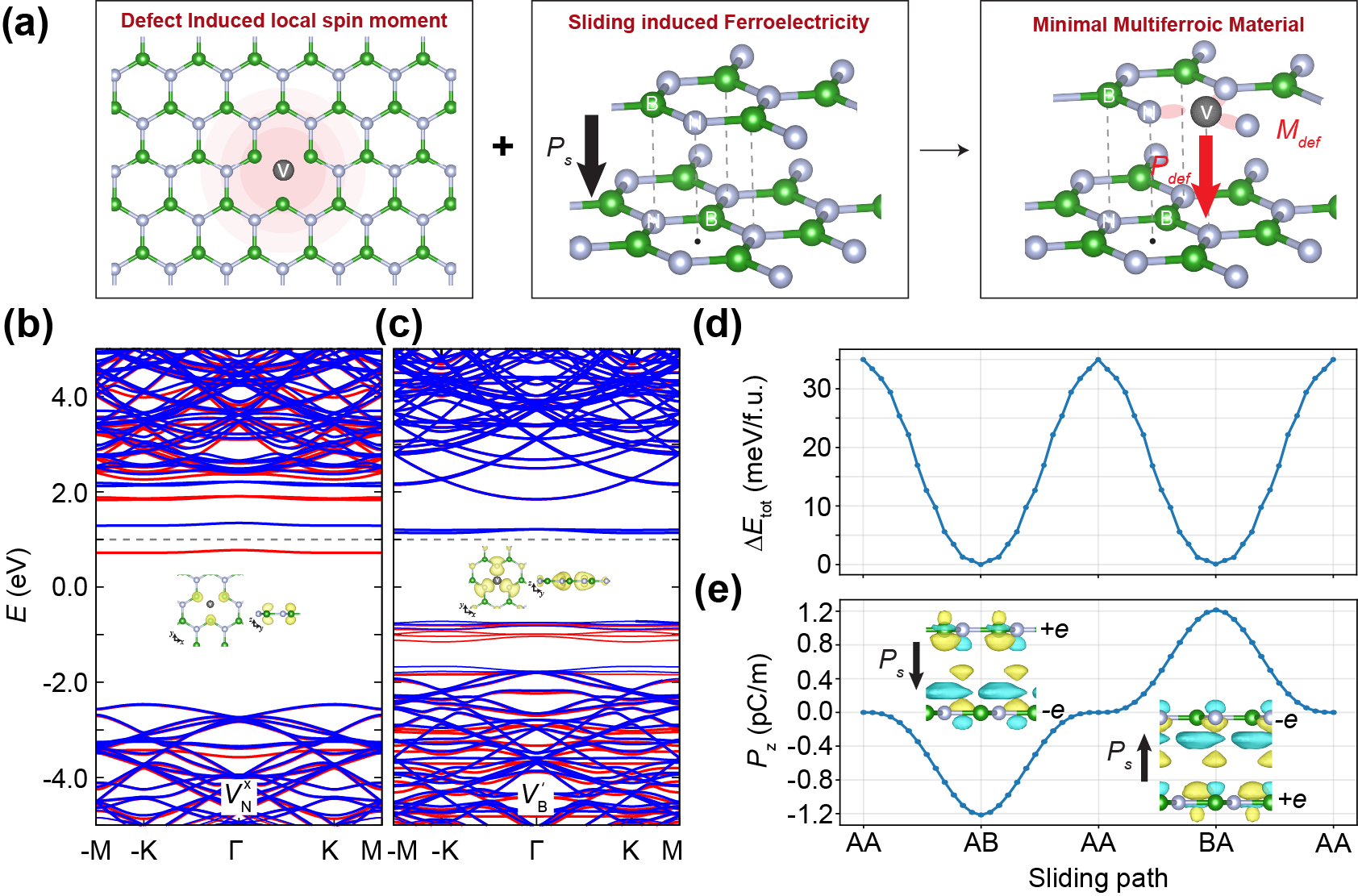}
\caption{
\textbf{Defect-enabled sliding multiferroic-like platform in bilayer hBN.}
(a) Design concept for combining defect-induced local spin moments with sliding ferroelectricity in hBN (B: green, N: light gray, and vacancy: dark gray). 
(b and c) Spin-polarized band structures of monolayer hBN for (b) a nitrogen vacancy (V$_{\mathrm{N}}^{\times}$) and (c) a boron vacancy (V$_{\mathrm{B}}^{'}$). Red and blue bands denote the two spin channels (spin up: red, spin down: blue), and the Fermi level is denoted as gray dotted lines. Insets show the corresponding spin-density isosurfaces.
(d and e) (d) Relative total energy $\Delta E_{\mathrm{tot}}$ and (e) Out-of-plane Berry-phase polarization $P_z$ of pristine bilayer hBN along the interlayer sliding path $AA \rightarrow AB \rightarrow AA \rightarrow BA \rightarrow AA$. Insets in (e) illustrate the charge distribution ($\rho_{\rm{bilayer}}-\rho_{\rm{monolayer},\rm{top}}-\rho_{\rm{monolayer},\rm{bottom}}$) of AB and BA registries. Yellow and cyan charge isosurfaces denote electron accumulation and depletion, respectively.
}
  \label{fig1}
\end{figure*}

In this Letter, we demonstrate that atomic defects convert the sliding registry $Q_s$ of bilayer hBN into a local magnetic control parameter using a phenomenological model and first-principles calculations. A single vacancy lifts the AB-BA equivalence, while interlayer vacancy pairs exhibit defect-dependent ferrimagnetic (FiM) or antiferromagnetic (AFM) coupling. For the AFM V$_{\mathrm{N}}^{\times}$--V$_{\mathrm{N}}^{\times}$ pair, a finite out-of-plane electric field selects the polar registry and modulates the magnitude of the layer-resolved magnetic order parameter.
Crucially, by modulating the layer-resolved N\'eel order via an out-of-plane electric field, we bypass the fundamental scarcity of 2D multiferroics and establish a robust spintronic readout paradigm. This framework provides versatile nonvolatile magnetoelectric control across a broad class of polar sliding bilayers and moir\'e heterostructures. 

\blue{\emph{Single defect and sliding ferroelectricity in hBN}}.--- We first examine the effect of a single defect in bilayer hBN. As illustrated in Fig.~\ref{fig1}(a), point vacancies in the honeycomb lattice of hBN generate localized spin magnetic moments, and bilayer hBN can exhibit out-of-plane polarization depending on the stacking registry. We propose that combining these two components can realize a minimal multiferroic platform in which defect-localized magnetic moments are embedded in a switchable sliding-ferroelectric host.

To analyze the behavior of these two components, we examine their electronic structures using first-principles calculations. Detailed methods are summarized in the Supplemental Material~\cite{SM}. We calculated the electronic structures of $5\times5$ monolayer hBN supercells containing a single N or B vacancy, as shown in Figs.~\ref{fig1}(b) and \ref{fig1}(c), respectively. All calculations examine V$_{\mathrm N}^{\times}$ and V$_{\mathrm B}^{\prime}$ vacancy charge states that can be simultaneously stabilized within a physically accessible Fermi-level range in  hBN~\cite{Weston2018PRB,Maciaszek2022PRM,Maciaszek2026npj2DMater}. Details related to the choice of charge state for defects are provided in the Supplemental Material~\cite{SM}. Each vacancy generates a six-orbital defect manifold formed from one $p_z$ orbital and one $sp^2$ dangling-bond orbital on each of the three neighboring atoms~\cite{Weston2018PRB,Ivady2020}. For $V_{\mathrm{N}}^{\times}$, the three electrons from the neighboring B atoms doubly occupy the lowest defect orbital and singly occupy the next, giving an $S=1/2$ ground state~\cite{Weston2018PRB}. For $V_{\mathrm{B}}^{'}$, total ten defect electrons from the three neighboring N atoms supplying nine electrons with additional electron from charged defect fill doubly occupied four of the six defect orbitals and singly occupied two in-plane $sp^2$-derived orbitals by parallel-spin electrons, yielding an $S=1$ ground state~\cite{Ivady2020}. The corresponding spin densities are localized on the three nearest-neighbor atoms, as shown in the insets of Figs.~\ref{fig1}(b) and \ref{fig1}(c).
These localized defect states carry magnetic moments of approximately $\pm2\,\mu_{\mathrm B}$ for V$_{\mathrm B}^{'}$ and $\pm1\,\mu_{\mathrm B}$ for V$_{\mathrm N}^{\times}$, where the sign indicates the spin-polarization direction. 


To examine the sliding-ferroelectric degree of freedom, the relative total-energy and polarization profiles of pristine hBN bilayer along the stacking path are summarized in Figs.~\ref{fig1}(d) and \ref{fig1}(e), respectively~\cite{Li2017ACSNano}. The energy profile contains two degenerate minima corresponding to the AB and BA registries, with the higher-energy nonpolar AA registry and the calculated switching barrier is approximately 35 meV/f.u.
. Berry-phase polarization calculations further show that the AB and BA registries have polarizations of exactly equal magnitude and opposite direction: $P_z(\mathrm{AB}) \approx -1.208\ \mathrm{pC/m}$ and
$P_z(\mathrm{BA}) \approx +1.208\ \mathrm{pC/m}$. We now proceed to determine whether these two components (sliding ferroelectricity and defect magnetism) merely coexist or are coupled through the stacking-dependent local environment.


\begin{figure}[t] \centering \includegraphics[width=\columnwidth]{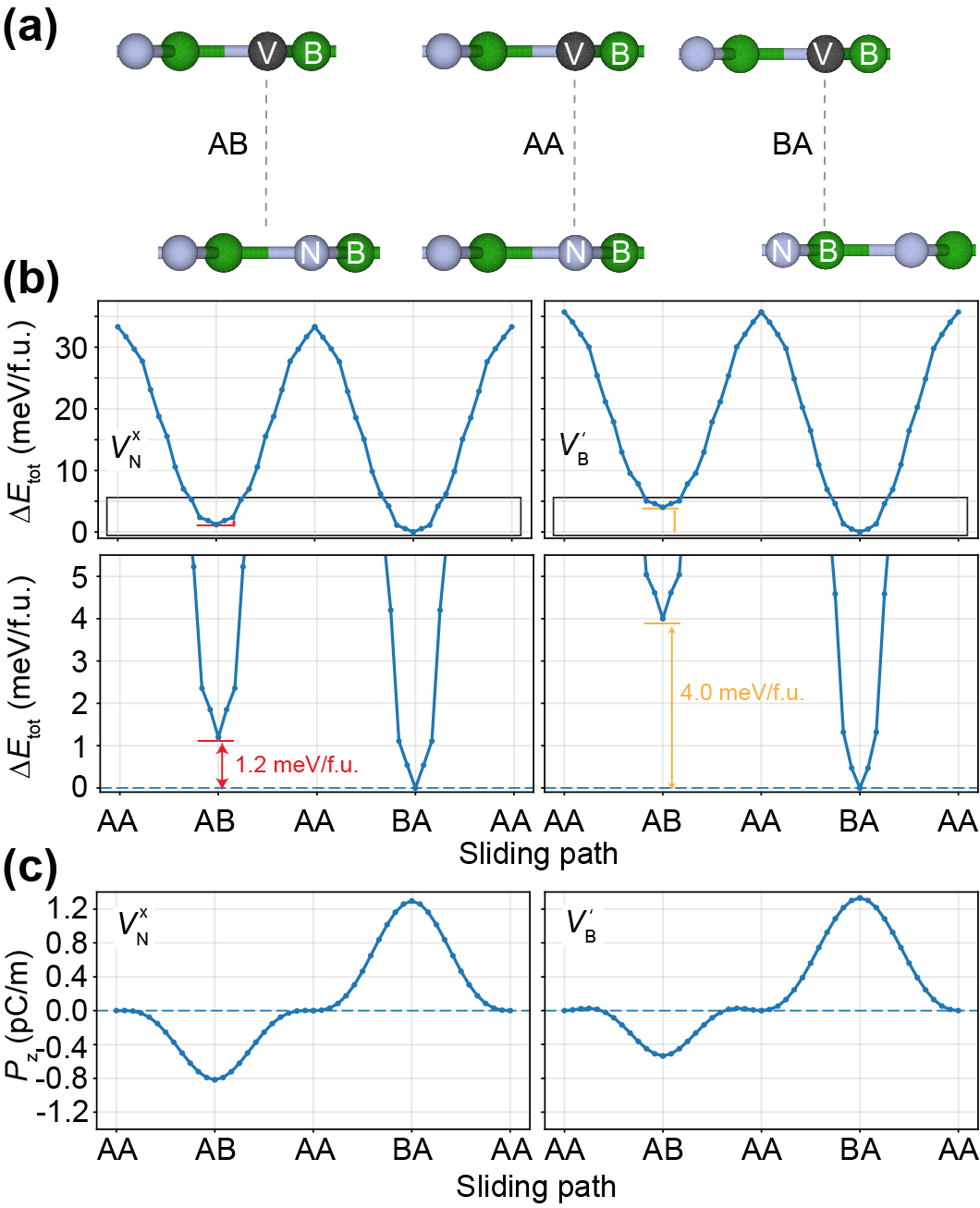} \caption{\textbf{Vacancy-induced inequivalence of AB and BA registries in sliding-ferroelectric bilayer hBN.} (a) Side-view atomic structures of bilayer hBN with a fixed top-layer N vacancy at the BA, AA, and AB stacking registries. The dashed line marks the lateral position of the vacancy site. (b and c) (b) Overall and zoom-in relative total energy $\Delta E_{\mathrm{tot}}$ and (c) out-of-plane Berry-phase polarization $P_z$ for bilayer hBN containing a nitrogen vacancy ($V_{\mathrm{N}}$, left) or a boron vacancy ($V_{\mathrm{B}}$, right) along the interlayer sliding path $AA \rightarrow AB \rightarrow AA \rightarrow BA \rightarrow AA$. Energies and polarizations are referenced to the lowest-energy and centrosymmetric configuration along each path.} \label{fig2} \end{figure}

\blue{\emph{Vacancy-selected ferroelectric registry in bilayer hBN}}.--- Having established the two ingredients, vacancy-localized magnetic moments and sliding ferroelectricity, we next examine whether the vacancy degree of freedom in bilayer hBN is independent of the sliding-ferroelectric coordinate or energetically coupled to the stacking registry. 
If the vacancy merely coexists with the sliding-ferroelectric background without coupling to the local registry, this AB-BA equivalence should remain intact. As illustrated in Fig.~\ref{fig2}(a), the local vertical environment of a top-layer vacancy depends on both the stacking registry and the vacancy species. For V$_{\mathrm{N}}$, the vacant N site lies above a hollow site in the AB registry, whereas it  lies above a bottom-layer B atom in the BA registry. In contrast, for V$_{\mathrm{B}}$, the vacant B site lies above a bottom-layer N atom in the AB registry and above a hollow site in the BA registry. Thus, for either vacancy
species, the AB and BA configurations expose the defect to distinct local interlayer environments. 

Consistent with this local inequivalence, the DFT energy profiles in Fig.~\ref{fig2}(b) show that one of the two ferroelectric partner states becomes energetically preferred. To describe vacancies in either layer, we define the defect-layer variable $\eta_{\mathrm{def}}=+1$ ($-1$) for a vacancy in the top (bottom) layer. For the top-layer configurations shown in Fig.~\ref{fig2}, corresponding to $\eta_{\mathrm{def}}=+1$, the calculated signed AB--BA energy differences are $E^{V_{\mathrm{N}}^{\times}}(\mathrm{BA})-E^{V_{\mathrm{N}}^{\times}}(\mathrm{AB})\approx-1.2~\mathrm{meV/f.u.}$ and $E^{V_{\mathrm{B}}^{\prime}}(\mathrm{BA})-E^{V_{\mathrm{B}}^{\prime}}(\mathrm{AB})\approx-4.0~\mathrm{meV/f.u.}$. 
Thus, the negative splitting energies for $\eta_{\mathrm{def}}=+1$ indicate that the BA registry is energetically preferred for a top-layer vacancy, whereas transferring the vacancy to the bottom layer reverses the sign of the splitting and selects the AB registry. The splitting for V$_{\mathrm{B}}^{\prime}$ is approximately $3.3$ times larger than that for V$_{\mathrm{N}}^{\times}$, showing that V$_{\mathrm{B}}^{\prime}$ couples more strongly to the stacking registry. 
For the polarization profiles shown in Fig.~\ref{fig2}(c), we decompose the calculated zero-field polarization for each vacancy species as $P_z(\eta_{\mathrm{def}})=P_{z,\mathrm{odd}}+\eta_{\mathrm{def}}P_{z,\mathrm{even}}$, where $P_{z,\mathrm{odd}}=\left[P_z(\mathrm{BA},\eta_{\mathrm{def}})-P_z(\mathrm{AB},\eta_{\mathrm{def}})\right]/2$ and $P_{z,\mathrm{even}}=\left[P_z(\mathrm{BA},\eta_{\mathrm{def}})+P_z(\mathrm{AB},\eta_{\mathrm{def}})\right]/(2\eta_{\mathrm{def}})$. Here, $P_{z,\mathrm{odd}}$ is the registry-odd sliding-ferroelectric polarization, whereas $P_{z,\mathrm{even}}$ is the registry-even polarization offset arising from the asymmetric distribution of defects between the two layers. The latter changes sign when the vacancy is transferred from one layer to the other. In the pristine limit, these quantities are $P_{z,\mathrm{odd}}^{0}=1.208$~pC/m and $P_{z,\mathrm{even}}^{0}=0$, respectively, as shown in Fig.~\ref{fig1}(e). At the defect density corresponding to one vacancy per $5\times5$ supercell, V$_{\mathrm{N}}^{\times}$ and V$_{\mathrm{B}}^{\prime}$ produce registry-even polarization offsets of $P_{z,\mathrm{even}}^{V_{\mathrm{N}}^{\times}}\approx0.240$~pC/m and $P_{z,\mathrm{even}}^{V_{\mathrm{B}}^{\prime}}\approx0.398$~pC/m, respectively. The corresponding registry-odd sliding components are $P_{z,\mathrm{odd}}^{V_{\mathrm{N}}^{\times}}\approx1.055$~pC/m and $P_{z,\mathrm{odd}}^{V_{\mathrm{B}}^{\prime}}\approx0.933$~pC/m. Thus, the magnitude of the polarization offset is approximately $1.7$ times larger for V$_{\mathrm{B}}^{\prime}$ than for V$_{\mathrm{N}}^{\times}$, whereas the magnitude of the registry-odd component is approximately $1.1$ times larger for V$_{\mathrm{N}}^{\times}$. Moreover, V$_{\mathrm{B}}^{\prime}$ produces both a larger AB--BA energy splitting and a larger polarization offset than V$_{\mathrm{N}}^{\times}$, but the corresponding enhancement factors are different. The energetic registry preference and the defect-induced polar redistribution therefore are not directly proportional.

To describe the structural, polar, and magnetic properties within a common framework, we introduce a single parent free-energy functional and progressively specialize it to the degrees of freedom active in each calculation. We define $Q_s$ as a continuous sliding parameter with $Q_s=+1$ for BA and $Q_s=-1$ for AB, $d$ as the distance between two defect sites, and the ternary defect-species variable $\tau_{\mathrm{def}}\in\{+1,0,-1\}$ for V$_{\mathrm{B}}^{\prime}$, pristine, and V$_{\mathrm{N}}^{\times}$ configurations, respectively. For $\tau_{\mathrm{def}}=0$, the value of $\eta_{\mathrm{def}}$ is immaterial because all single-defect contributions vanish. The unified free energy with external electric field along $z$-direction $\mathcal{E}_z$ is written as
\begin{align}
    & \mathcal{F}(Q_s, \eta_{\mathrm{def}}, \tau_{\mathrm{def}} ;\mathcal{E}_z) = \mathcal{F}_{\mathrm {pri}}(Q_s;\mathcal{E}_z) + \nonumber \\ 
    &\mathcal{F}_{\mathrm{1def}}(Q_s,\eta_{\mathrm{def}}, \tau_{\mathrm{def}};\mathcal{E}_z) + \mathcal{F}_{\mathrm{1def}}(Q_s,\bar{\eta}_{\mathrm{def}}, \bar{\tau}_{\mathrm{def}};\mathcal{E}_z) + \nonumber \\
    &\mathcal{F}_{\mathrm {int}}(Q_s, \eta_{\mathrm{def}}, \bar{\eta}_{\mathrm{def}}, \tau_{\mathrm{def}}, \bar{\tau}_{\mathrm{def}}, d;\mathcal{E}_z),
    \label{eq:parent_free_energy}
\end{align}
where $\mathcal{F}_{\mathrm{pri}}$, $\mathcal{F}_{\mathrm{1def}}$, and $\mathcal{F}_{\mathrm{int}}$ indicate the free-energy contributions from pristine bilayer hBN, a single vacancy, and vacancy--vacancy interactions, respectively. The pristine bilayer hBN contribution with sliding dependent polarization has $\mathcal{F}_{\mathrm{pri}}(Q_s;\mathcal{E}_z)=\sum_{ij} c_{ij}Q^i\mathcal{E}_z^j = c_{20}Q_s^2+c_{40}Q_s^4-c_{11}Q_s\mathcal{E}_z + c_{02}\mathcal{E}_z^2+\cdots$. We can fit our result in Figs.~\ref{fig1}(d) and (e) into $\mathcal{F}_{\mathrm{pri}}$ that $c_{20}=-88.63 \mathrm{meV/f.u.}$, $c_{40}=55.43 \mathrm{meV/f.u.}$, and $c_{11}=1.143 \mathrm{pC/m}$. In the limit of $d \rightarrow \infty$, $\mathcal{F}_{\mathrm {int}}$ will be negligible. Here, all calculations are performed at $\mathbf{B}=0$, so magnetic-field contributions are omitted below.

We first develop the single-defect part of Eq.~\eqref{eq:parent_free_energy}, as shown in Fig.~\ref{fig2}(a). Under all symmetry operations of AA-stacked bilayer hBN ($D_{3h}$)
, the layer-preserving operations $\{E,2C_{3z},3\sigma_v\}$ leave both $Q_s$, $\eta_{\mathrm{def}}$, and $\mathcal{E}_z$ unchanged, whereas the layer-exchanging operations $\{3C_2',\sigma_h,2S_3\}$ reverse them all. Consequently, the products $\eta_{\mathrm{def}}Q_s$, $\mathcal{E}_z Q_s$, and $\mathcal{E}_z\eta_{\mathrm{def}}$ are invariant under every symmetry operation of the AA reference structure and constitute the lowest-order couplings between the defect layer, the polar registry, and the external field. The dependence of these couplings on defect type can be represented by the lowest-order quadratic polynomial in $\tau_{\mathrm{def}}$. In particular, $\tau_{\mathrm{def}}^2$ distinguishes defective configurations from the pristine configuration, whereas $\tau_{\mathrm{def}}$ distinguishes V$_{\mathrm{B}}^{\prime}$ from V$_{\mathrm{N}}^{\times}$. Using the DFT energy splittings and polarization components obtained above, the corresponding minimal single-defect contribution for the present $5\times5$ supercell is
\begin{align}
&\mathcal{F}_{\mathrm{1def}}(Q_s,\eta_{\mathrm{def}},\tau_{\mathrm{def}};\mathcal{E}_z) = \sum_{ijkl} c_{ijkl} Q_s^i \eta_{\mathrm{def}}^j \tau_{\mathrm{def}}^k \mathcal{E}_z^l
 \nonumber \\
&=c_{1110} Q_s \eta_{\mathrm{def}} \tau_{\mathrm{def}} + c_{1120} Q_s \eta_{\mathrm{def}} \tau_{\mathrm{def}}^2 \nonumber \\
&+c_{1011}Q_s \tau_{\mathrm {def}} \mathcal{E}_z + c_{1021} Q_s \tau_{\mathrm{def}}^2 \mathcal{E}_z \nonumber \\
&+c_{0111}\eta_{\mathrm{def}} \tau_{\mathrm{def}} \mathcal{E}_z + c_{0121}\eta_{\mathrm{def}} \tau_{\mathrm{def}}^2 \mathcal{E}_z + \cdots
\label{eq:structural_polar_free_energy}
\end{align}
The first two terms describe the coupling among the sliding registry, defect type, and defect layer. The next two terms represent the defect-induced correction to the registry-odd sliding polarization, whereas the last two terms describe the registry-even polarization offset arising from the asymmetric distribution of defects between
the two layers. A simultaneous fit to the calculated total-energy and polarization in Fig.~\ref{fig2}(b) and (c) gives
$c_{1110}=-0.70 \mathrm{meV/f.u.}$, $c_{1120}=-1.30 \mathrm{meV/f.u.}$, $c_{1011}=-0.061 \mathrm{pC/m}$, $c_{1021}=-0.214 \mathrm{pC/m}$, $c_{0111}=0.079 \mathrm{pC/m}$, and $c_{0121} = 0.319 \mathrm{pC/m}$.

\begin{figure}[t]
\centering
\includegraphics[width=\columnwidth]{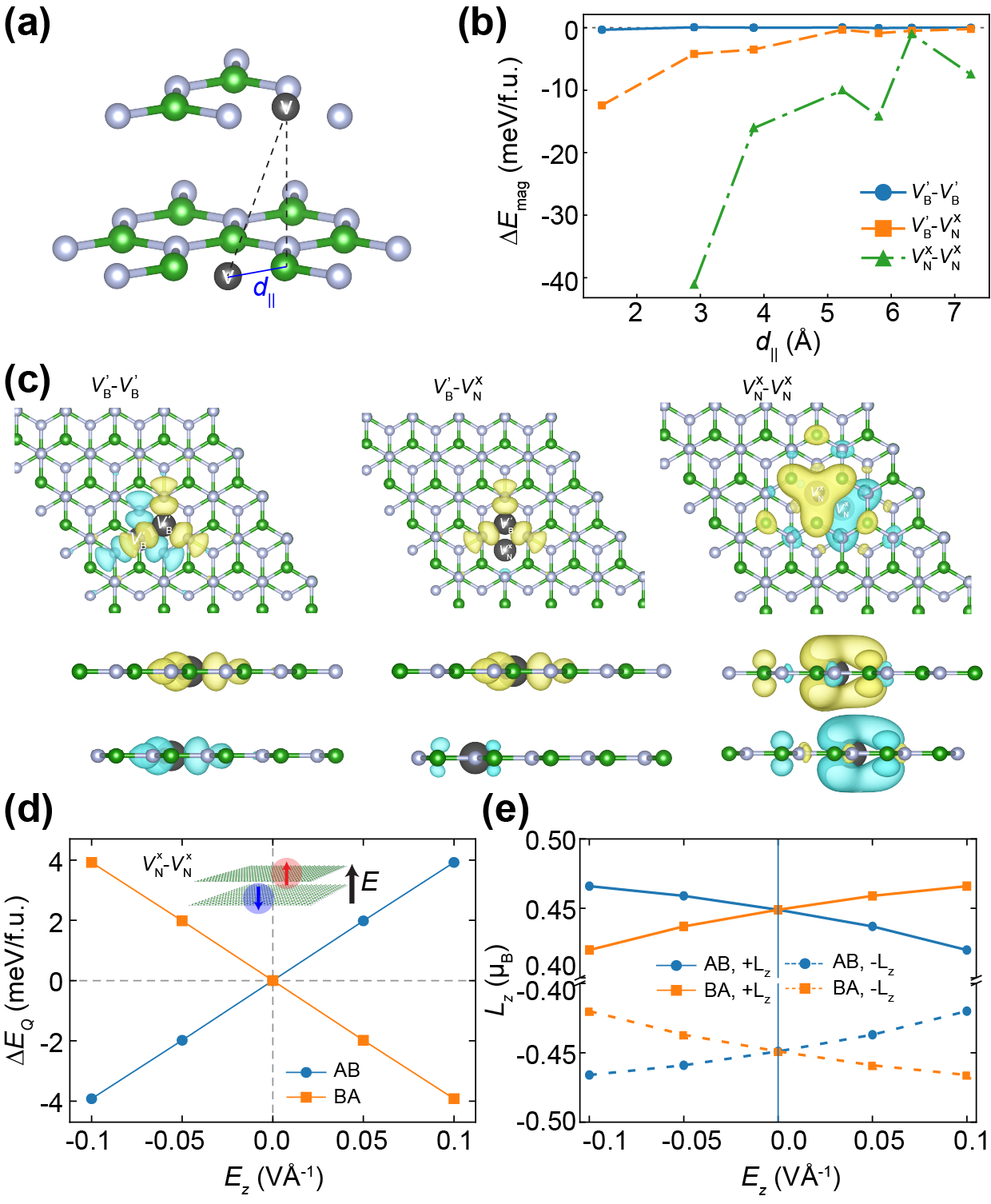}
\caption{\textbf{Interlayer vacancy-pair magnetism and electric-field-tunable antiferromagnetic order in sliding-ferroelectric bilayer hBN.} (a) Schematic side view of bilayer hBN containing one vacancy in each layer. The dashed line defines the in-plane defect--defect separation $d_{\parallel}$. (b) Magnetic energy difference $\Delta E_{\mathrm{mag}}=E_{\mathrm{AP}}-E_{\mathrm{P}}$ for $V_{\mathrm{B}}^{\prime}$--$V_{\mathrm{B}}^{\prime}$, $V_{\mathrm{B}}^{\prime}$--$V_{\mathrm{N}}^{\times}$, and $V_{\mathrm{N}}^{\times}$--$V_{\mathrm{N}}^{\times}$ pairs as a function of $d_{\parallel}$. (c) Spin-density isosurfaces for representative vacancy-pair configurations. (d) Relative energies $\widetilde{E}_{Q}$ of the AB and BA registries after subtraction of the common field-dependent contribution. (e) Layer-resolved N\'eel order, $L_z=(M_{\mathrm{top},z}-M_{\mathrm{bottom},z})/2$, showing a registry-dependent change in $L_z$ of up to $10.9\%$.}
\label{fig3}
\end{figure}

\blue{\emph{Electric-field control of a defect-coupled antiferromagnetic texture}}.--- To realize the magnetoelectric properties of interest, the magnetic properties must be controllable by an electric field. We first performed fully relativistic calculations to test whether sliding changes the orientation of an isolated defect-localized spin moment. For both vacancy types, however, we find a weak preference for out-of-plane magnetic anisotropy in both the AB and BA registries~\cite{SM}. Sliding therefore does not reverse the preferred orientation of a single-vacancy magnetic moment, motivating us to consider interlayer vacancy pairs.

We consider bilayer hBN containing one vacancy in each layer, as shown in Fig.~\ref{fig3}(a). For the interlayer vacancy pairs $V_{\mathrm{B}}^{\prime}$--$V_{\mathrm{B}}^{\prime}$, $V_{\mathrm{B}}^{\prime}$--$V_{\mathrm{N}}^{\times}$, and $V_{\mathrm{N}}^{\times}$--$V_{\mathrm{N}}^{\times}$, we compare the parallel and antiparallel spin configurations and define $\Delta E_{\mathrm{mag}}\equiv E_{\mathrm{AP}}-E_{\mathrm{P}}$. Under this definition, $\Delta E_{\mathrm{mag}}<0$ indicates that the antiparallel configuration is favored, whereas $\Delta E_{\mathrm{mag}}>0$ indicates that the parallel configuration is favored.
Figure~\ref{fig3}(b) shows that the magnetic coupling depends strongly on the vacancy composition and separation. For the $V_{\mathrm{B}}^{\prime}$--$V_{\mathrm{B}}^{\prime}$ pair, the antiparallel configuration is favored by only approximately $0.34$~meV at the shortest separation. At larger separations, the energy splitting is generally below $0.07$~meV and changes sign with distance, indicating that the two defect moments are nearly magnetically decoupled rather than robustly antiferromagnetically coupled. In contrast, the mixed $V_{\mathrm{B}}^{\prime}$--$V_{\mathrm{N}}^{\times}$ pair favors the antiparallel configuration over the full range of separations considered in the BA registry. Because the two defect moments have unequal magnitudes and therefore do not fully compensate, this antiparallel configuration is FiM. Its stabilization is approximately $12.4$~meV at the shortest separation and decreases to approximately $0.2$~meV at the largest separation considered, with a weak nonmonotonic modulation at intermediate distances. The $V_{\mathrm{N}}^{\times}$--$V_{\mathrm{N}}^{\times}$ pair, by contrast, forms an AFM state stabilized by approximately $2$--$40$~meV. Its net magnetization is $\langle\mathbf{M}\rangle=0$, while the real-space magnetization density remains finite and has opposite signs in the two layers.

These trends follow from the distinct defect orbitals. V$_{\mathrm N}^{\times}$ hosts a singly occupied predominantly $p_z$ state, so two V$_{\mathrm N}^{\times}$ defects form a half-filled two-site system in which virtual hopping favors AFM alignment. In contrast, the predominantly in-plane $sp^2$ states of V$_{\mathrm B}^{'}$ have weaker interlayer overlap, yielding much weaker exchange. Mixed V${\mathrm B}^{'}$–V$_{\mathrm N}^{\times}$ pairs couple unequal $S=1$ and $S=1/2$ moments antiparallel, producing a FiM state.
The corresponding effective models and microscopic interpretations are discussed in the Supplemental Material~\cite{SM}. 
Because electrostatic gating can shift the Fermi level and thereby alter the stable charge states of V$_{\mathrm{B}}$ and V$_{\mathrm{N}}$, it may provide an additional route to tune the local moments and magnetic ground states of vacancy pairs~\cite{Weston2018PRB,Maciaszek2022PRM,Gale2023NanoLett}.

We focus on the AFM V$_{\mathrm{N}}^{\times}$--V$_{\mathrm{N}}^{\times}$ pair as the clearest route for controlling a compensated layer-resolved magnetic texture. We define an order parameter as ${L_z}=(\mathbf{M}_{z,\mathrm{top}}-\mathbf{M}_{z,\mathrm{bottom}})/2$.
To determine its response to an out-of-plane electric field, we compare the AB and BA registries for the two time-reversed configurations, $L_z=\pm|L_z|$. At $E_z=0$, all four states are degenerate within numerical accuracy. A finite $E_z$ lifts the AB--BA degeneracy approximately linearly, reaching $3.92$~meV at $|E_z|=0.10~\mathrm{V}/\text{\AA}$, and reversing the field reverses the preferred registry [Fig.~\ref{fig3}(d)]. At each fixed registry, however, the two time-reversed states remain degenerate within $0.3~\mu$eV, as required by time-reversal symmetry in the absence of a magnetic field. Accordingly, $\mathcal{F}(L_z)=\mathcal{F}(-L_z)$. Using the same symmetry classification of the AA reference structure introduced above, $Q_s$ and $E_z$ are both odd under layer-exchanging operations, whereas $L_z^2$ is invariant under all spatial operations and time reversal. Thus, $Q_sE_zL_z^2$ is the lowest-order field-linear invariant that distinguishes the two polar registries and modulates the AFM amplitude without lifting the degeneracy between the time-reversed N\'eel states.

Specializing Eq.~\eqref{eq:parent_free_energy} to an interlayer V$_{\mathrm{N}}^{\times}$--V$_{\mathrm{N}}^{\times}$ pair in AB/BA registry, the minimal defect-free magnetic free energy can be written as
\begin{align}
&\mathcal{F}_{\mathrm{int}}(Q_s,d,L_z;\mathcal{E}_z)
=\sum_{ikl}c_{ikl}(d)Q_s^iL_z^k\mathcal{E}_z^l
\nonumber\\
&=\left[c_{020}(d)+c_{220}(d)Q_s^2\right]L_z^2
+c_{040}(d)L_z^4
-c_{121}(d)Q_sL_z^2\mathcal{E}_z+\cdots .
\label{eq:pair_free_energy}
\end{align}
Here, $c_{020}(d)+c_{220}(d)Q_s^2$ is the zero-field quadratic coefficient of the AFM order parameter, while $c_{040}(d)>0$ stabilizes the ordered state. The field-linear term $-c_{121}(d)Q_sL_z^2E_z$ changes sign upon reversing either the registry or the electric field, thereby selecting between AB and BA while preserving the degeneracy between $L_z$. For $Q_s=\pm1$, the calculated splitting of $3.92$~meV at $|E_z|=0.10~\mathrm{V}/\text{\AA}$ gives $c_{121}(d=2.9\,\text{\AA})L_z^2=0.0196\,e\,\text{\AA}$. Minimization with respect to $L_z$ gives
\begin{align}
L_z^2=-\frac{c_{020}(d)+c_{220}(d)Q_s^2}{2c_{040}(d)}+\frac{c_{121}(d)}{2c_{040}(d)}Q_sE_z+\cdots, 
\label{eq:neel_amplitude_field} 
\end{align}
capturing the increase of $|L_z|$ for $Q_sE_z>0$ for a positive $c_{121}(d)$. As shown in Fig.~\ref{fig3}(e), the field-favored registry exhibits the larger defect-localized moment, yielding a registry-dependent contrast of  $\approx 10.9\%$ at an applied field of $|E_z|=0.10~\mathrm{V}/\text{\AA}$. Fully relativistic noncollinear calculations show that the electric field modulates the magnitude rather than the orientation of the N\'eel order, as the nearly antiparallel layer moments remain compensated. This field-induced modulation of $L_z$ could provide physical and technological significance. First, it demonstrates a purely electronic control of magnetism in a non-magnetic parent system with fully compensated emergent AFM. The mechanism is completely driven by sliding-mediated interlayer charge transfer and orbital hybridization, allowing the amplitude of the N\'eel vector to be continuously dialed up or down. 
Second, because the sign of the field-induced change in the N\'eel-order magnitude, $\frac{\partial |L_z|}{\partial E_z}$ in \cite{SM}, is strictly determined by the stacking registry, it establishes a definitive spintronic readout pathway for sliding ferroelectricity. The local stacking state can be non-destructively read by monitoring the asymmetric change in local spin-density magnitudes or spin-transport under a small probing electric field. 


\blue{\emph{Conclusion}}.--- In summary, we demonstrate a defect-engineered coupling between sliding ferroelectricity and magnetism in bilayer hBN. While isolated vacancies break the polar AB-BA degeneracy, interlayer pairs provide sublattice-selective magnetic ground states yielding either FiM (V$_{\mathrm{B}}^{\prime}$--V$_{\mathrm{N}}^{\times}$) or AFM (V$_{\mathrm{N}}^{\times}$--V$_{\mathrm{N}}^{\times}$) configurations. 
The $10.9\%$ registry contrast under an out-of-plane electric field provides direct evidence of sliding-mediated orbital magnetoelectricity, offering a purely electrostatic pathway to tune a compensated N\'eel amplitude without external magnetic fields or relativistic spin-orbit coupling. By showing that the defect sublattices dictate the interlayer exchange, this work delivers a novel framework for engineering multiferroic functionality in nonmagnetic vdWs systems, paving the way to a clean foundation for 2D slidetronic logic and memory architecture. This framework generalizes to polar bilayers, moiré domains, and TMD heterostructures where localized moments couple to switchable stacking registries~\cite{woods2021charge,Wang2022NatNano}. 


\blue{\emph{Acknowledgment}}.--- This work was supported by the U.S. Department of Energy, Office of Science, Basic Energy Sciences, under Award No. DE-SC0024942. Computational support was provided by the National Energy Research Scientific Computing Center (NERSC), a U.S. Department of Energy, Office of Science User Facility located at Lawrence Berkeley National Laboratory, operated under Contract No. DE-AC02-05CH11231.

\blue{\emph{Data Availablity}}.--- The data supporting the findings of this study are available in the Dryad
Digital Repository~\cite{Kim2026Dryad}.

\bibliography{reference}

\section*{End Matter}
\begin{figure}[ht]
\centering
\includegraphics[width=\columnwidth]{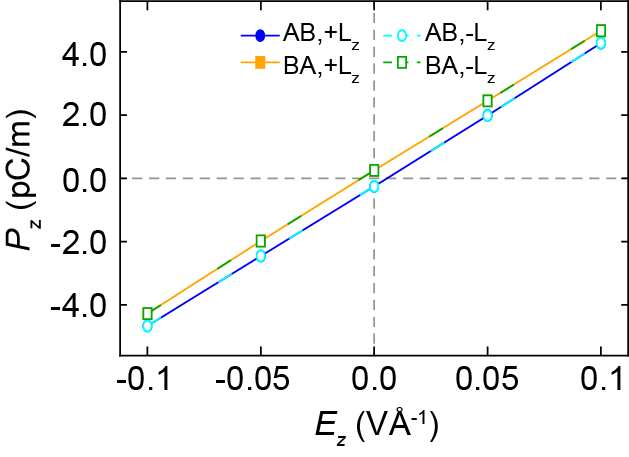}
\caption{Out-of-plane Berry-phase polarization $P_z$ for the AFM (V$_\mathrm{N}$-V$_\mathrm{N}$) bilayer hBN depending on E-field}
\label{fig4}
\end{figure}
To verify the self-consistency of this framework, we examine $P_z$ vs. electric-field for both the AB and BA registries and $\pm L_z$ N\'eel directions. As shown in Fig.~\ref{fig4} (in End matter), the electric polarizations for the $+L_z$ and $-L_z$ configurations are completely degenerate and map onto identical linear trajectories for a given stacking registry, which confirms that $P_z$ also is invariant under a global reversal of the layer N\'eel vector. From a phenomenological standpoint, this directly validates the structural formulation of Eq.~\eqref{eq:pair_free_energy}, demonstrating that the magnetoelectric coupling is consistent with leading order $Q_sE_zL^2_z$ term, while linear terms in $L_z$ are strictly forbidden by time-reversal symmetry. Moreover, the identical slopes indicate that the linear dielectric susceptibility is highly robust against sliding-induced changes in the stacking registry.

\clearpage
\onecolumngrid

\begin{center}
{\Large Supplementary Materials for}\\[0.5em]
{\large \textbf{Defect-Controlled Multiferroicity via Stacking Control in Nonmagnetic van der Waals Bilayers}}\\[1em]
Bumseop Kim, Sayed Ali Akbar Ghorashi, Andrew M. Rappe
\end{center}

\vspace{1em}

\setcounter{section}{0}
\setcounter{equation}{0}
\setcounter{figure}{0}
\setcounter{table}{0}

\renewcommand{\thesection}{S\arabic{section}}
\renewcommand{\thesubsection}{\thesection.\Alph{subsection}}
\renewcommand{\theequation}{S\arabic{equation}}
\renewcommand{\thefigure}{S\arabic{figure}}
\renewcommand{\thetable}{S\arabic{table}}

\renewcommand{\theHsection}{S\arabic{section}}
\renewcommand{\theHsubsection}{S\arabic{section}.\Alph{subsection}}
\renewcommand{\theHequation}{S\arabic{equation}}
\renewcommand{\theHfigure}{S\arabic{figure}}
\renewcommand{\theHtable}{S\arabic{table}}

\section{Computational Methods}
\label{sec:computational_methods}

First-principles calculations were performed within spin-polarized density functional theory (DFT) using the plane-wave pseudopotential method implemented in the QUANTUM ESPRESSO package~\cite{Giannozzi2009JPCM,Giannozzi2017JPCM}. Exchange-correlation effects were treated within the generalized gradient approximation using the Perdew--Burke--Ernzerhof functional~\cite{Perdew1996PRL}, and the ionic cores were represented by norm-conserving pseudopotentials. Long-range dispersion interactions between the hBN layers were described using the DFT-D3 method~\cite{Grimme2010JCP}. Kinetic-energy cutoffs of 60 and 240~Ry were used for the wave functions and charge density, respectively. The Brillouin zone of the $5\times5$ bilayer hBN supercell was sampled using a $5\times5\times1$ Monkhorst--Pack grid~\cite{Monkhorst1976PRB}. A vacuum region of 20~\AA{} was introduced along the out-of-plane direction, and the electronic self-consistency threshold was set to $1.0\times10^{-8}$~Ry.

Most structural, polarization, and vacancy-pair exchange calculations were performed using collinear spin-polarized DFT with scalar-relativistic pseudopotentials. Magnetic-anisotropy calculations were performed self-consistently using fully relativistic pseudopotentials and a noncollinear spinor Hamiltonian including spin--orbit coupling. The finite-field calculations used to resolve the orientation and layer-projected amplitudes of the $V_{\mathrm N}^{\times}$--$V_{\mathrm N}^{\times}$ N\'eel order were performed within the same fully relativistic noncollinear framework. All calculations were performed at zero external magnetic field.

The interlayer sliding paths were constructed by rigidly translating one hBN layer relative to the other while keeping the intralayer atomic coordinates and interlayer separation fixed. A neutral nitrogen vacancy $V_{\mathrm N}^{\times}$ was modeled by removing one N atom from a charge-neutral supercell. A negatively charged boron vacancy $V_{\mathrm B}^{\prime}$ was modeled by removing one B atom and adding one electron. The total supercell charges were therefore $-2$, $-1$, and $0$ for the $V_{\mathrm B}^{\prime}$--$V_{\mathrm B}^{\prime}$, $V_{\mathrm B}^{\prime}$--$V_{\mathrm N}^{\times}$, and $V_{\mathrm N}^{\times}$--$V_{\mathrm N}^{\times}$ pairs, respectively. Charged periodic supercells were treated using the uniform compensating background implemented in QUANTUM ESPRESSO. For single-defect calculations, the vacancy was placed in either the top or bottom layer while the stacking registry was varied; for vacancy-pair calculations, one vacancy was placed in each layer.

For all vacancy pairs, parallel (P) and antiparallel (AP) solutions were obtained by initializing the defect-localized moments with parallel and antiparallel orientations. For identical-defect pairs, the P and AP solutions correspond to FM and compensated AFM arrangements, respectively. For the mixed $V_{\mathrm B}^{\prime}$--$V_{\mathrm N}^{\times}$ pair, the AP solution is ferrimagnetic because the two defect moments have unequal magnitudes. The magnetic energy difference is defined throughout as
\begin{equation}
\Delta E_{\mathrm{mag}}=E_{\mathrm{AP}}-E_{\mathrm{P}}.
\label{eq:sm_delta_emag}
\end{equation}

The out-of-plane electric polarization was evaluated using the Berry-phase formulation of the modern theory of polarization~\cite{KingSmith1993PRB}. The three-dimensional polarization obtained from the periodic supercell was converted to a two-dimensional polarization according to $P_z^{\mathrm{2D}}=L_z^{\mathrm{cell}}P_z^{\mathrm{3D}}$, where $L_z^{\mathrm{cell}}$ is the out-of-plane supercell length and should not be confused with the layer-resolved N\'eel-order parameter introduced below. For charged $V_{\mathrm B}^{\prime}$-containing cells, only fixed-charge polarization differences evaluated using the same supercell, compensating background, and Berry-phase branch are compared. Out-of-plane electric fields ranging from $-0.10$ to $+0.10~\mathrm{V}/\text{\AA}$ were applied using a sawtooth electrostatic potential together with a dipole correction.

\subsection{Selection of Vacancy Charge States}
Previous hybrid-functional calculations for bulk hBN reported the transition levels $\varepsilon(0/-1)=1.48$~eV and $\varepsilon(-1/-2)=4.90$~eV for $V_{\mathrm B}$, and $\varepsilon(+1/0)=3.48$~eV and $\varepsilon(0/-1)=4.59$~eV for $V_{\mathrm N}$, where the Fermi level is referenced to the valence-band maximum and the calculated band gap is 5.94~eV \cite{Weston2018PRB}. The corresponding thermodynamically stable vacancy states are summarized in Table~\ref{tab:vacancy_charge_states}.
\begin{table}[t]
\centering
\caption{Thermodynamically stable charge states of isolated boron and nitrogen vacancies as a function of the Fermi level. The Fermi level is referenced to the valence-band maximum. At each transition level, the two adjacent charge states are degenerate.}
\label{tab:vacancy_charge_states}
\begin{tabular}{ccc}
\hline\hline
$E_F-E_{\mathrm{VBM}}$ (eV)
& Stable $V_{\mathrm B}$ state
& Stable $V_{\mathrm N}$ state \\
\hline
$0.00$--$1.48$
& $V_{\mathrm B}^{\times}$
& $V_{\mathrm N}^{\bullet}$ \\
$1.48$--$3.48$
& $V_{\mathrm B}^{\prime}$
& $V_{\mathrm N}^{\bullet}$ \\
$3.48$--$4.59$
& $V_{\mathrm B}^{\prime}$
& $V_{\mathrm N}^{\times}$ \\
$4.59$--$4.90$
& $V_{\mathrm B}^{\prime}$
& $V_{\mathrm N}^{\prime}$ \\
$4.90$--$5.94$
& $V_{\mathrm B}^{\prime\prime}$
& $V_{\mathrm N}^{\prime}$ \\
\hline\hline
\end{tabular}
\end{table}
In a real hBN sample, the Fermi level is determined collectively by the concentrations of all native defects, impurities, free carriers, and external charge reservoirs rather than by the vacancies explicitly included in a particular simulation cell. The formation energy of a defect $D$ in the charge state $\alpha$ is expressed as
\begin{equation}
\Delta H_f(D^{\alpha};E_F)=E_{\mathrm{tot}}(D^{\alpha})-E_{\mathrm{tot}}(\mathrm{host})-\sum_i n_i\mu_i+q_{\alpha}(E_F+E_{\mathrm{VBM}})+E_{\mathrm{corr}},\label{eq:defect_formation_energy}
\end{equation}
where $n_i$ is the number of atoms of species $i$ added to the supercell, $\mu_i$ is the corresponding chemical potential, $q_{\alpha}$ is the numerical charge associated with the charge state $\alpha$, and $E_{\mathrm{corr}}$ is the finite-size electrostatic correction for a charged supercell. The equilibrium concentration of each defect state at temperature $T$ is then given by
\begin{equation}
N(D^{\alpha};E_F,T)=N_{D,\mathrm{sites}}g_{D,\alpha}\exp\left[-\frac{\Delta H_f(D^{\alpha};E_F)}{k_{\mathrm B}T}\right],
\label{eq:defect_concentration}
\end{equation}
where $N_{D,\mathrm{sites}}$ is the density of available defect sites and $g_{D,\alpha}$ is the configurational and electronic degeneracy factor. The equilibrium Fermi level is obtained by solving Eq.~\eqref{eq:defect_concentration} self-consistently with the global charge-neutrality condition
\begin{equation}
p(E_F)-n(E_F)+\sum_D\sum_{\alpha}q_{\alpha}N(D^{\alpha};E_F,T)+Q_{\mathrm{ext}}=0,
\label{eq:charge_neutrality}
\end{equation}
where $p$ and $n$ are the free-hole and free-electron concentrations, respectively, and $Q_{\mathrm{ext}}$ represents charge supplied by substrates, electrodes, interfaces, or other external reservoirs. Because of the large band gap of hBN, the equilibrium Fermi level is generally governed predominantly by compensation between positively and negatively charged defects rather than by free carriers. Defect-thermodynamic calculations including carbon- and hydrogen-related impurities have applied Eqs.~\eqref{eq:defect_concentration} and \eqref{eq:charge_neutrality} to determine the Fermi level self-consistently~\cite{Maciaszek2022PRM}. Under N-rich conditions at a representative growth temperature of 1500~K, an equilibrium value of $E_F-E_{\mathrm{VBM}}=3.63$~eV was obtained for carbon- and hydrogen-containing hBN, with the Fermi-level position governed primarily by compensation between the positively charged carbon donor $C_{\mathrm B}^{\bullet}$ and the negatively charged vacancy--hydrogen complex $(V_{\mathrm B}-2\mathrm H)^{\prime}$~\cite{Maciaszek2026npj2DMater}.  Alternatively, under carbon-rich conditions intermediate between the N-rich and N-poor limits, the presence of oxygen raises the equilibrium Fermi level to approximately $3.6$~eV at 1600~K, likewise stabilizing $V_{\mathrm B}^{\prime}$ and $V_{\mathrm N}^{\times}$ simultaneously~\cite{Maciaszek2022PRM}. This value lies within the $3.48$--$4.59$~eV interval in Table~\ref{tab:vacancy_charge_states}, for which $V_{\mathrm B}^{\prime}$ and $V_{\mathrm N}^{\times}$ are simultaneously thermodynamically stable. We therefore assume an N-rich, impurity-containing hBN environment and adopt $E_F-E_{\mathrm{VBM}}\simeq3.6$~eV as a physically motivated reference for selecting the vacancy charge states. Accordingly, all calculations in this work use $V_{\mathrm B}^{\prime}$ and $V_{\mathrm N}^{\times}$. The compensating impurities are not included explicitly; they represent spatially remote reservoirs that establish the physical Fermi level. In the periodic first-principles calculations, the net charges of $V_{\mathrm B}^{\prime}$-containing supercells are compensated by the uniform background described in Sec.~\ref{sec:computational_methods}. This construction preserves the spin-active configurations of both vacancy species and enables a consistent comparison of $V_{\mathrm B}^{\prime}$--$V_{\mathrm B}^{\prime}$, $V_{\mathrm B}^{\prime}$--$V_{\mathrm N}^{\times}$, and $V_{\mathrm N}^{\times}$--$V_{\mathrm N}^{\times}$ pairs.

\section{Single-Defect Magnetic Anisotropy in the AB and BA Registries}

To examine whether the sliding registry controls the spin orientation of an isolated vacancy, fully relativistic calculations including spin--orbit coupling were performed for in-plane and out-of-plane spin orientations. The magnetic anisotropy energy is defined as
\begin{equation}
\Delta E_{\mathrm{MAE}}=E_{\mathrm{in}}-E_{\mathrm{out}}.
\label{eq:sm_mae}
\end{equation}
As summarized in Table~\ref{tab:single_defect_mae}, the out-of-plane orientation is lower in energy for both $V_{\mathrm B}^{\prime}$ and $V_{\mathrm N}^{\times}$ in the AB and BA registries. The anisotropy is weak, ranging from approximately $0.1$ to $4.9~\mu\mathrm{eV}$, and the preferred orientation does not reverse under AB--BA sliding.

\begin{table}[h]
\caption{Fully relativistic total energies for in-plane and out-of-plane spin orientations of isolated $V_{\mathrm B}^{\prime}$ and $V_{\mathrm N}^{\times}$ defects. Positive $\Delta E_{\mathrm{MAE}}$ indicates an out-of-plane preferred orientation.}
\label{tab:single_defect_mae}
\begin{ruledtabular}
\begin{tabular}{ccccc}
Defect & Registry & $E_{\mathrm{in}}$ (Ry) & $E_{\mathrm{out}}$ (Ry) & $\Delta E_{\mathrm{MAE}}$ ($\mu$eV) \\
\hline
$V_{\mathrm B}^{\prime}$ & AB & $-1333.862420720$ & $-1333.862420940$ & $2.99$ \\
$V_{\mathrm B}^{\prime}$ & BA & $-1333.862496810$ & $-1333.862497170$ & $4.90$ \\
$V_{\mathrm N}^{\times}$ & AB & $-1319.169218750$ & $-1319.169218760$ & $0.14$ \\
$V_{\mathrm N}^{\times}$ & BA & $-1319.171616320$ & $-1319.171616510$ & $2.59$ \\
\end{tabular}
\end{ruledtabular}
\end{table}

Because these anisotropy energies are on the microelectronvolt scale, their precise values are sensitive to total-energy convergence. They should therefore be interpreted primarily as showing a weak out-of-plane preference that does not reverse under AB--BA sliding.

\section{Microscopic Interpretation of Vacancy-Pair Magnetism}

The three vacancy pairs exhibit qualitatively different magnetic regimes. Throughout this section, $\Delta E_{\mathrm{mag}}=E_{\mathrm{AP}}-E_{\mathrm{P}}$ follows Eq.~\eqref{eq:sm_delta_emag}. The $V_{\mathrm N}^{\times}$--$V_{\mathrm N}^{\times}$ pair has a compensated AP ground state with an AFM stabilization of approximately $2$--$40$~meV. In the BA registry, the mixed $V_{\mathrm B}^{\prime}$--$V_{\mathrm N}^{\times}$ pair favors an AP state by approximately $0.2$--$12.4$~meV; because the two moments are unequal, this state is FiM. By contrast, the $V_{\mathrm B}^{\prime}$--$V_{\mathrm B}^{\prime}$ pair retains sizable local moments but exhibits only sub-meV P--AP splittings, with no robust FM or AFM preference over most separations. The following effective models rationalize these trends without implying a unique microscopic decomposition of every exchange contribution.

\subsection{Antiferromagnetic kinetic exchange between two neutral nitrogen vacancies}

A neutral $V_{\mathrm N}^{\times}$ hosts one unpaired electron in a defect state distributed over the three neighboring B atoms with predominantly out-of-plane $p_z$ character. Two such defects can therefore be represented minimally by one approximately singly occupied effective orbital on each vacancy center. The resulting half-filled two-site Hubbard Hamiltonian is
\begin{equation}
H=-t\sum_{\sigma}\left(c_{1\sigma}^{\dagger}c_{2\sigma}+c_{2\sigma}^{\dagger}c_{1\sigma}\right)
+U\sum_{i=1,2}n_{i\uparrow}n_{i\downarrow},
\label{eq:sm_hubbard_vn}
\end{equation}
where $t$ is the hopping amplitude between the two effective defect orbitals and $U$ is the energy cost of double occupation.

At half filling, virtual hopping lowers the singlet-like antiparallel sector but not the triplet-like parallel sector. To second order in $t/U$,
\begin{equation}
\Delta E_{S}^{(2)}=-\frac{4t^2}{U},
\qquad
\Delta E_{T}^{(2)}=0,
\label{eq:sm_second_order_energy}
\end{equation}
and projection onto the singly occupied subspace gives
\begin{equation}
H_{\mathrm{eff}}
=
J\left(\mathbf S_1\cdot\mathbf S_2-\frac14\right),
\qquad
J=\frac{4t^2}{U}>0.
\label{eq:sm_afm_exchange_vn}
\end{equation}
Accordingly,
\begin{equation}
\Delta E_{\mathrm{mag}}
=
E_{\mathrm{AP}}-E_{\mathrm{P}}
\sim
-\frac{4t(d_{\parallel})^2}{U},
\label{eq:sm_exchange_distance}
\end{equation}
which approaches zero as the interdefect hopping decreases with increasing in-plane separation. This minimal kinetic-exchange picture is consistent with the calculated AFM sign and overall decay of the $V_{\mathrm N}^{\times}$--$V_{\mathrm N}^{\times}$ energy splitting. The AP solution obtained in spin-polarized DFT is a broken-symmetry representation of the singlet-like coupled state rather than an exact many-body singlet eigenstate.

\subsection{Weak exchange between two negatively charged boron vacancies}

A $V_{\mathrm B}^{\prime}$ defect is an effective $S=1$ center derived primarily from the N-centered, in-plane $e^{\prime}$ dangling-bond manifold of the isolated defect. In the pair calculations, the P solution has a total magnetization close to $4~\mu_{\mathrm B}$, whereas the AP solution has nearly zero net magnetization. Their absolute magnetizations remain nearly equal, demonstrating that the two local moments survive in both configurations rather than being quenched.

The localized and predominantly in-plane character of the $V_{\mathrm B}^{\prime}$ orbitals is consistent with reduced effective overlap between opposite layers. Correspondingly, the P--AP energy difference is only approximately $0.34$~meV at the shortest BA separation and generally below $0.07$~meV at larger separations. The small calculated sign variations in this sub-meV regime do not establish a robust sequence of FM--AFM transitions. The supported conclusion is instead that the two $V_{\mathrm B}^{\prime}$ moments are nearly magnetically decoupled over most of the separations examined.

Because the present calculations do not independently determine defect-resolved hopping matrices, direct-exchange integrals, or multi-orbital interaction parameters, the weak $V_{\mathrm B}^{\prime}$--$V_{\mathrm B}^{\prime}$ coupling is not assigned to a unique exchange channel. A quantitative decomposition would require, for example, localized Wannier orbitals, constrained-spin calculations, or an explicit mapping onto a multi-orbital interacting Hamiltonian.

\subsection{Antiparallel ferrimagnetic coupling in the mixed pair}

The mixed $V_{\mathrm B}^{\prime}$--$V_{\mathrm N}^{\times}$ pair combines effective local moments corresponding approximately to $S=1$ and $S=1/2$. In the BA registry, the P solutions retain total magnetizations of approximately $2.7$--$2.9~\mu_{\mathrm B}$, while the lower-energy AP solutions retain an uncompensated moment of approximately $1~\mu_{\mathrm B}$. The finite absolute magnetization in both solutions confirms that the two defect moments remain active. Their unequal magnitudes prevent complete cancellation in the AP configuration, which is therefore FiM rather than a compensated AFM state.

The AP stabilization decreases overall from approximately $12.4$~meV at the shortest separation to approximately $0.2$~meV at the largest separation examined, indicating a short-ranged interaction between the inequivalent defect manifolds. Its weak nonmonotonic distance dependence also shows that the coupling depends on detailed orbital registry rather than on the scalar separation alone. The calculations establish the survival of two unequal moments and their preferred antiparallel alignment in the BA registry, but they do not uniquely separate hybridization-mediated exchange from accompanying charge redistribution. We therefore interpret the mixed-pair result as an effective antiferromagnetic coupling between unequal local moments, producing a FiM state.

\section{Finite-Field Analysis of the Antiferromagnetic $V_{\mathrm N}^{\times}$--$V_{\mathrm N}^{\times}$ Pair}
\label{sec:finite_field_nn}

The AB and BA registries were calculated for both globally spin-reversed AFM configurations at $E_z=-0.10$, $-0.05$, $0$, $+0.05$, and $+0.10~\mathrm{V}/\text{\AA}$. The electric field was introduced using the sawtooth-potential and dipole-correction procedure described in Sec.~\ref{sec:computational_methods}, and each registry and spin configuration was converged independently.

\subsection{Layer-resolved order parameter}

The layer moments were obtained from the out-of-plane magnetization density $m_z(\mathbf r)$ by dividing the supercell at the midpoint $z_0$ between the two layers:
\begin{align}
M_{z,\mathrm{top}}&=\int_{z>z_0}m_z(\mathbf r)\,d^3r,\\
M_{z,\mathrm{bottom}}&=\int_{z<z_0}m_z(\mathbf r)\,d^3r.
\label{eq:sm_layer_moments}
\end{align}
The layer-resolved N\'eel order is
\begin{equation}
L_z=\frac{M_{z,\mathrm{top}}-M_{z,\mathrm{bottom}}}{2}.
\label{eq:sm_layer_neel_order}
\end{equation}
Under time reversal, $L_z\rightarrow-L_z$, whereas $Q_s$ and $E_z$ remain unchanged.

\subsection{Field-dependent energies and time-reversal degeneracy}

For each registry $Q\in\{\mathrm{AB},\mathrm{BA}\}$, we define the spin-averaged energy and time-reversal splitting as
\begin{equation}
E_Q(E_z)=\frac{E_Q(+L_z;E_z)+E_Q(-L_z;E_z)}{2},
\qquad
\delta_{\mathrm{TR}}^Q(E_z)=E_Q(+L_z;E_z)-E_Q(-L_z;E_z).
\label{eq:sm_spin_averaged_energy}
\end{equation}
The calculated registry and time-reversal splittings are summarized in Table~\ref{tab:finite_field_energies}. The BA--AB splitting changes sign when the electric field is reversed, whereas the largest residual time-reversal splitting is only $0.272~\mu\mathrm{eV}$, which is within the numerical precision of the independently converged calculations.

\begin{table}[h]
\caption{Field-dependent energy difference between the BA and AB registries and absolute energy splittings between the two globally spin-reversed AFM configurations.}
\label{tab:finite_field_energies}
\begin{ruledtabular}
\begin{tabular}{cccc}
$E_z$ ($\mathrm{V}/\text{\AA}$)
&
$E_{\mathrm{BA}}-E_{\mathrm{AB}}$ (meV)
&
$|\delta_{\mathrm{TR}}^{\mathrm{AB}}|$ ($\mu$eV)
&
$|\delta_{\mathrm{TR}}^{\mathrm{BA}}|$ ($\mu$eV)
\\
\hline
$-0.100$ & $+3.921297$ & $0.000$ & $0.000$ \\
$-0.050$ & $+1.981329$ & $0.136$ & $0.000$ \\
$0.000$  & $0.000000$  & $0.000$ & $0.000$ \\
$+0.050$ & $-1.981397$ & $0.000$ & $0.272$ \\
$+0.100$ & $-3.921297$ & $0.000$ & $0.000$ \\
\end{tabular}
\end{ruledtabular}
\end{table}

\subsection{Cancellation of the leading single-defect registry coupling}

For an isolated vacancy, the leading zero-field registry coupling has the form $c_{\mathrm N}\eta_{\mathrm{def}}Q_s$. In the identical interlayer $V_{\mathrm N}^{\times}$--$V_{\mathrm N}^{\times}$ pair, one vacancy occupies each layer, so that $\eta_1=+1$ and $\eta_2=-1$. The leading single-defect contributions therefore cancel:
\begin{equation}
\sum_{i=1}^{2}c_{\mathrm N}\eta_iQ_s
=
c_{\mathrm N}Q_s(\eta_1+\eta_2)
=
0.
\label{eq:sm_single_defect_cancellation}
\end{equation}
This cancellation is consistent with the symmetry-required zero-field degeneracy of the AB and BA registries for the identical pair.

\subsection{Symmetry-constrained finite-field free energy}

The same AA-reference symmetry classification used in the main text gives $Q_s\rightarrow-Q_s$ and $E_z\rightarrow-E_z$ under layer-exchanging operations, whereas $L_z^2$ is invariant under spatial operations and time reversal. Thus, $Q_sE_zL_z^2$ is the lowest-order field-linear invariant that distinguishes the two registries and modifies the AFM amplitude without selecting the sign of $L_z$.

For fitting the total finite-field energies, we retain the direct polar coupling and the leading field-even correction in addition to the minimal magnetic free energy given in the main text:
\begin{align}
\mathcal F(Q_s,d,L_z;E_z)
={}&F_{\mathrm c}(E_z)-AP_{\mathrm{pair}}Q_sE_z \nonumber\\
&+\left[a(d)-\gamma(d)Q_sE_z+\zeta(d)E_z^2\right]L_z^2
+bL_z^4+\cdots .
\label{eq:sm_finite_field_free_energy}
\end{align}
Here, $F_{\mathrm c}(E_z)$ contains contributions common to the two registries, $A$ is the in-plane supercell area, $P_{\mathrm{pair}}$ is the registry-odd two-dimensional polarization of the pair system, and $\gamma(d)$ describes the registry-dependent field modulation of the AFM amplitude. The coefficient $\zeta(d)$ captures the leading field-even correction omitted from the field-linear minimal expression in the main text.

Because only even powers of $L_z$ occur,
\begin{equation}
\mathcal F(Q_s,+L_z;E_z)=\mathcal F(Q_s,-L_z;E_z),
\label{eq:sm_time_reversal_degeneracy}
\end{equation}
consistent with Table~\ref{tab:finite_field_energies}. Within the ordered phase, minimization gives
\begin{equation}
|L_z|^2
=
-\frac{a-\gamma Q_sE_z+\zeta E_z^2}{2b}
=
L_0^2+\frac{\gamma}{2b}Q_sE_z-\frac{\zeta}{2b}E_z^2,
\qquad
L_0^2=-\frac{a}{2b},
\label{eq:sm_neel_minimum}
\end{equation}
where the implicit distance dependence is omitted for compactness. The $Q_sE_zL_z^2$ invariant therefore changes the equilibrium magnitude of the N\'eel order without lifting the degeneracy between $\pm L_z$.

\subsection{Fit of the field-dependent registry energy}

To remove the field-dependent contribution common to the AB and BA registries, we define
\begin{equation}
\widetilde E_Q(E_z)
=
E_Q(E_z)
-
\frac{E_{\mathrm{AB}}(E_z)+E_{\mathrm{BA}}(E_z)}{2}.
\label{eq:sm_relative_energy}
\end{equation}
After minimizing over $L_z$, the leading registry-dependent contribution is
\begin{equation}
\widetilde E_Q(E_z)
=
-p_{\mathrm{eff}}Q_sE_z+\mathcal O(E_z^3),
\qquad
p_{\mathrm{eff}}=AP_{\mathrm{pair}}+\gamma L_0^2.
\label{eq:sm_effective_dipole}
\end{equation}
Thus, the fitted registry energy contains both the direct polar contribution and the magnetic renormalization associated with the equilibrium AFM amplitude.

A linear fit gives
\begin{equation}
E_{\mathrm{BA}}(E_z)-E_{\mathrm{AB}}(E_z)
=
-39.296~\mathrm{meV}\,
\frac{E_z}{1~\mathrm{V}/\text{\AA}},
\label{eq:sm_registry_energy_fit}
\end{equation}
with $R^2=0.99998$. Since $E_{\mathrm{BA}}-E_{\mathrm{AB}}=-2p_{\mathrm{eff}}E_z$, the fitted slope corresponds to
\begin{equation}
p_{\mathrm{eff}}=0.01965~e\,\text{\AA}.
\label{eq:sm_effective_dipole_value}
\end{equation}
The total-energy data determine only the combined coefficient $p_{\mathrm{eff}}$. Separating $AP_{\mathrm{pair}}$ from $\gamma L_0^2$ would require an independent determination of the pair polarization contribution or calculations at constrained $L_z$.

\subsection{Fit of the layer-resolved N\'eel-order magnitude}

The magnitudes obtained from Eqs.~\eqref{eq:sm_layer_moments} and \eqref{eq:sm_layer_neel_order}, averaged over the two time-reversed configurations, are listed in Table~\ref{tab:finite_field_neel_order}.

\begin{table}[h]
\caption{Magnitude of the layer-resolved N\'eel order for the AB and BA registries as a function of the applied electric field.}
\label{tab:finite_field_neel_order}
\begin{ruledtabular}
\begin{tabular}{ccc}
$E_z$ ($\mathrm{V}/\text{\AA}$)
&
$|L_z|_{\mathrm{AB}}$ ($\mu_{\mathrm B}$)
&
$|L_z|_{\mathrm{BA}}$ ($\mu_{\mathrm B}$)
\\
\hline
$-0.100$ & $0.466186$ & $0.420416$ \\
$-0.050$ & $0.459388$ & $0.436674$ \\
$0.000$  & $0.449755$ & $0.449313$ \\
$+0.050$ & $0.436866$ & $0.458817$ \\
$+0.100$ & $0.420560$ & $0.465364$ \\
\end{tabular}
\end{ruledtabular}
\end{table}

A simultaneous fit of both registries to Eq.~\eqref{eq:sm_neel_minimum} gives
\begin{align}
|L_z|^2
={}&0.20208~\mu_{\mathrm B}^2
+0.20056~\mu_{\mathrm B}^2Q_s\frac{E_z}{1~\mathrm{V}/\text{\AA}}
\nonumber\\
&-0.51977~\mu_{\mathrm B}^2
\left(\frac{E_z}{1~\mathrm{V}/\text{\AA}}\right)^2,
\label{eq:sm_neel_fit}
\end{align}
with $R^2=0.99974$. Comparison with Eq.~\eqref{eq:sm_neel_minimum} yields
\begin{equation}
L_0=0.4495~\mu_{\mathrm B},
\qquad
\frac{\gamma}{2b}=0.20056~\mu_{\mathrm B}^2\frac{\text{\AA}}{\mathrm V},
\qquad
\frac{\zeta}{2b}=0.51977~\mu_{\mathrm B}^2\frac{\text{\AA}^2}{\mathrm V^2}.
\label{eq:sm_landau_fit_parameters}
\end{equation}
The positive value of $\gamma/(2b)$ shows that $|L_z|$ increases for $Q_sE_z>0$. Defining the registry contrast as
\begin{equation}
C_L(E_z)
=
\frac{|L_z|_{\mathrm{fav}}-|L_z|_{\mathrm{unfav}}}
{|L_z|_{\mathrm{unfav}}}
\times100\%,
\label{eq:sm_registry_contrast}
\end{equation}
the maximum calculated contrast is approximately $10.9\%$ at $|E_z|=0.10~\mathrm{V}/\text{\AA}$. The present data determine the ratios $\gamma/b$ and $\zeta/b$, but not the individual Landau coefficients.

\subsection{Polarization under global N\'eel reversal}

For each fixed registry and electric field, the Berry-phase polarizations of the $+L_z$ and $-L_z$ configurations are indistinguishable within numerical accuracy, as shown in Fig.~4 of the main text. Thus,
\begin{equation}
P_z(Q_s,+L_z;E_z)=P_z(Q_s,-L_z;E_z),
\label{eq:sm_polarization_time_reversal}
\end{equation}
consistent with the absence of terms odd in $L_z$. The nearly identical polarization slopes obtained for the two time-reversed states further show that the linear dielectric response does not distinguish the sign of the N\'eel order.

\end{document}